\pdfoutput=1
\documentclass[conference]{IEEEtran}
\IEEEoverridecommandlockouts

\usepackage{cite}
\usepackage{amsmath,amssymb}
\usepackage{graphicx}
\usepackage{booktabs}
\usepackage{tabularx}
\usepackage{multirow}
\usepackage{stfloats}
\usepackage{xcolor}
\usepackage{url}
\usepackage{xspace}
\graphicspath{{figures/}}

\definecolor{linkblue}{HTML}{1F4E9E}
\usepackage[colorlinks,linkcolor=linkblue,citecolor=linkblue,urlcolor=linkblue,
  pdfauthor={},pdftitle={},pdfsubject={},pdfkeywords={},pdfcreator={},pdfproducer={}]{hyperref}
\usepackage{refcount}
\newcommand{\figlink}[5]{\put(#1,#2){\hyperlink{\getrefbykeydefault{#5}{anchor}{}}{\rule{0pt}{#4bp}\hspace{#3bp}}}}

\newcommand{\rev}[1]{#1}
\newcommand{\sys}{Weave\xspace}
\newcommand{\x}{$\times$\xspace}

\begin{document}

\title{\sys: Routing-Adaptive Runtime Warp Scheduling for MoE Megakernels}

\author{%
\IEEEauthorblockN{%
Ziyu Huang$^{*1}$, Yangjie Zhou$^{*2}$, Chenhao Zhu$^{1}$, Hongzhe Yan$^{1}$, Tianwei Qiao$^{1}$, Sicheng He$^{1}$, Zihan Liu$^{1}$,\\
Peng Yu$^{3}$, Shulai Zhang$^{1}$, Xingxun Tang$^{1}$, Xinhao Luo$^{1}$, Minyi Guo$^{1}$, Xiu Lin$^{4}$, Yinghao Yu$^{4}$,\\
Guodong Yang$^{4}$, Liping Zhang$^{4}$, Shixuan Sun$^{1}$, Qingyun Chen$^{5}$, Jingwen Leng$^{1}$}
\IEEEauthorblockA{%
$^{1}$Shanghai Jiao Tong University, Shanghai, China\quad
$^{2}$National University of Singapore, Singapore\\
$^{3}$Fudan University, Shanghai, China\;\;
$^{4}$Alibaba Group, China\;\;
$^{5}$Shanghai TokMach Technology Co., Ltd., Shanghai, China\\
huang\_ziyu@sjtu.edu.cn}
\thanks{$^{*}$Both authors contributed equally to this work.}}

\maketitle

\begin{abstract}
The Mixture-of-Experts (MoE) architecture is widely used in mainstream large language models. Under expert parallelism (EP), state-of-the-art systems fuse each MoE layer into a persistent megakernel for high performance. By the nature of EP, each GPU's communication and computation volumes vary with the routing result. However, prior work either fixes how GPU resources are divided between communication and computation at compile time (e.g., MegaMoE), or adapts to runtime dynamism only at coarse-grained SM granularity (e.g., mKernel), dedicating whole SMs to communication and leaving their Tensor Cores idle. At the finer warp granularity within each SM, static warp schedules can neither match routing-dependent loads nor prevent warps from idling while waiting for inputs or after finishing their assigned work. Therefore, we propose \sys, an MoE megakernel with three key designs. First, a \emph{spatial schedule} allocates resources between communication and computation at both SM and warp granularities to match each GPU's load. Second, a \emph{temporal schedule} switches warp roles as execution progresses, enlisting all warps at the start and end of a layer and selectively handing output scatter to idle communication warps. Third, a fused \emph{cost model} jointly selects both schedules per GPU at launch from runtime routing statistics. Across all evaluated platforms, \sys outperforms the strongest baselines: under EP8 over DeepGEMM-MegaMoE, it speeds up the MoE layer by 1.30\x geomean and up to 1.97\x on SM103, and by 1.24\x geomean and up to 1.72\x on SM90; under EP16 on CDNA3, it beats the fastest official RCCL+AITER baseline by \rev{1.44\x} geomean (up to \rev{2.87\x}; \rev{1.67\x} over vLLM).

\end{abstract}

\section{Introduction}\label{sec:intro}

\begin{figure}[t]
\centering
\includegraphics[width=0.92\columnwidth]{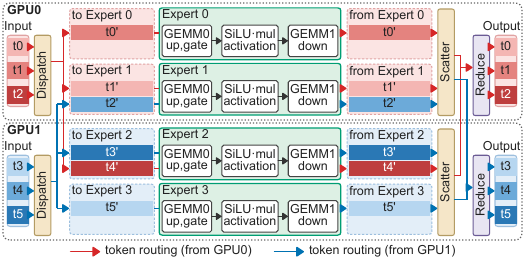}
\caption{Execution of one MoE layer under EP.}
\label{fig:moe}
\end{figure}

\rev{Mixture-of-Experts (MoE) architectures~\cite{gshard,switch,mixtral,deepseekv2,deepseekv3} are widely used in modern LLMs---such as DeepSeek-V4.1~\cite{deepseekv41}, Kimi K3.1~\cite{kimik31}, Qwen4~\cite{qwen4}, and GLM-5.3~\cite{glm53}---activating only a few of hundreds of experts per token. Under expert parallelism (EP), state-of-the-art systems fuse each MoE layer into one persistent megakernel~\cite{deepgemm,mpk,eventtensor}: as Fig.~\ref{fig:moe} shows, a layer dispatches tokens to GPUs holding their selected experts, runs GEMM0, activation, and GEMM1 per expert, scatters outputs back across GPUs, and reduces the weighted results locally.}

\rev{Under EP, however, token routing causes each GPU's communication and computation loads to diverge at runtime.} When a token selects multiple experts on one destination GPU, dispatch transfers its input once while each expert performs its own GEMMs, so deduplicated input communication and per-expert computation vary independently with routing.

\begin{figure}[t]
\centering
\includegraphics[width=0.92\columnwidth]{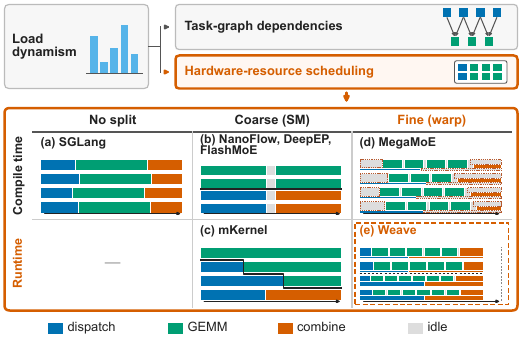}
\caption{Load dynamism (top) and hardware-resource scheduling across spatial granularity and decision time (bottom).}
\label{fig:patterns}\label{fig:levels}
\end{figure}

As Fig.~\ref{fig:levels} shows, routing dynamism affects both \emph{task-graph dependencies} and \emph{hardware-resource scheduling}. Existing megakernels already resolve runtime inter-operator tile readiness dynamically as inputs arrive~\cite{deepgemm,mpk,eventtensor,flashdmoe}.

\looseness=-1 In hardware-resource scheduling, however, prior schemes are coarse-grained or static (Fig.~\ref{fig:levels}). Without splitting, SGLang~\cite{sglang} serializes operators across kernels (Fig.~\ref{fig:levels}a). Coarse-grained schemes split whole SMs at compile time---NanoFlow~\cite{nanoflow}, DeepEP~\cite{deepep}, and FlashMoE~\cite{flashdmoe} (Fig.~\ref{fig:levels}b)---or at runtime, as in mKernel~\cite{mkernel} (Fig.~\ref{fig:levels}c), leaving communication SMs' Tensor Cores idle. At warp granularity, MegaMoE~\cite{deepgemm} shares each SM across communication and computation warps, but fixes warp schedules at compile time, preventing load adaptation and leaving warps idle (Fig.~\ref{fig:levels}d). Addressing both limitations requires scheduling warps within SMs at runtime (Fig.~\ref{fig:levels}e).

\looseness=-1 To address this challenge, we propose \sys, a routing-adaptive MoE megakernel with three key designs. First, a \emph{spatial schedule} partitions communication and computation warps at both SM and warp granularities, tuning intra-SM ratios and cross-SM mixes per GPU. Second, a \emph{temporal schedule} switches warp roles across phases, enlisting all warps in initial input pulls and final reduction while selectively offloading scatter to freed communication warps. Third, a fused \emph{cost model} jointly selects both schedules per GPU at launch from runtime routing statistics. On 8\x SM103, \sys beats DeepGEMM-MegaMoE by 1.30\x geomean (up to 1.97\x), FlashInfer by 1.40\x, and SGLang plus our mKernel, FlashMoE, and NanoFlow re-implementations by 3.26--5.00\x; on 8\x SM90 it beats MegaMoE by \rev{1.24}\x (up to \rev{1.72}\x) and FlashInfer by \rev{1.95\x}; on 16\x CDNA3 (EP16) it beats the fastest official RCCL+AITER baseline by \rev{1.44\x} (up to \rev{2.87\x}; \rev{1.67\x} over vLLM~\cite{vllm}). We will open-source our code soon.

\section{Background and Motivation}\label{sec:motiv}

\subsection{MoE Megakernel Execution}\label{sec:fixed}

MegaMoE executes an MoE layer with persistent thread blocks~\cite{persistent,streamk} active across operators. \rev{Fig.~\ref{fig:moe}'s \emph{dispatch} maps to peer-to-peer input \emph{pull}, followed after expert GEMMs by cross-GPU output \emph{scatter} and top-$k$ \emph{reduce}.} Within each SM on SM103, communication warps pull peer inputs, TMA warps load weights and activations, MMA warps issue Tensor Core instructions, and epilogue warps process and write back results. However, MegaMoE fixes both schedules at compile time: spatially, the communication-to-compute warp ratio is static; temporally, each warp remains locked to one role.

\subsection{Challenges}\label{sec:why}

\begin{figure*}[!b]
\centering
\includegraphics[width=0.94\textwidth]{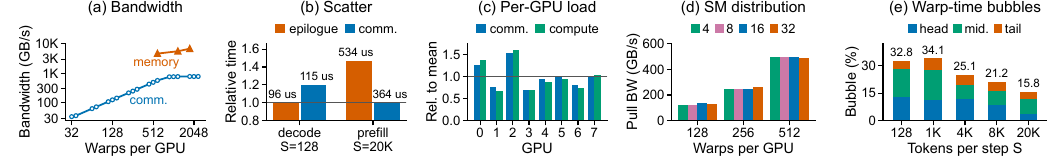}
\caption{\looseness=-1 Spatial and temporal scheduling factors on SM103 (EP8). (a) NVLink pull (4\,GiB) and epilogue TMEM-to-HBM (8\,GiB). (b) Layer latency by scatter path (Qwen3.5-122B). (c) Per-GPU comm./compute loads (MiniMax-M3, $S$=4K). (d) 4\,GiB pull across warps/SM. (e) MegaMoE warp-time bubbles (Qwen3.5-397B).}
\label{fig:warprole}
\end{figure*}

\begin{figure*}[t]
\centering
\includegraphics[width=\textwidth]{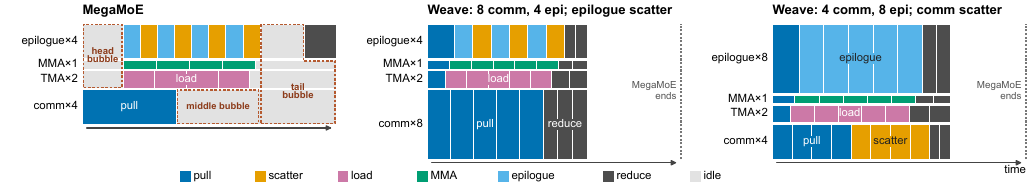}
\caption{MegaMoE vs.\ representative \sys schedules on SM103 (schematic).}
\label{fig:rolegraph}
\end{figure*}

\subsubsection{\rev{The spatial schedule must match the load}}

\looseness=-1 Figure~\ref{fig:warprole}c shows per-GPU demands for MiniMax-M3 ($S$=4K, EP8): dispatch inputs are deduplicated per destination GPU while GEMM work scales with token--expert pairs, yielding distinct per-GPU distributions.

\looseness=-1 \rev{Matching these demands requires tuning the warp ratio: communication bandwidth scales with total communication warps rather than how many SMs host them. On SM103, pull bandwidth saturates at 780~GB/s with 888 warps (Fig.~\ref{fig:warprole}a), and for a fixed total (128--512 warps), placing 4--32 warps/SM varies bandwidth by under 1\% (Fig.~\ref{fig:warprole}d). Meanwhile, epilogue throughput rises from 4.6 to 5.6 and 6.8~TB/s with 4, 8, and 12 epilogue warps/SM, governing GEMM accumulator drain and throughput.}

\subsubsection{\rev{Static temporal schedules} leave head, middle, and tail bubbles}

\rev{As Fig.~\ref{fig:rolegraph} (left) illustrates, MegaMoE incurs head, middle, and tail bubbles. Defining \emph{GPU warp-time} as total launched warps times layer duration on SM103, in the \emph{head} phase GEMM warps stall 24.3--36.2~$\mu$s awaiting inputs, wasting 11.0\%--12.7\% of warp-time at $S \le 1$K and 3.7\% at 20K.} In the \emph{middle} phase, communication warps idle once pulls finish---\rev{wasting 15.3\%--16.9\% at $S \le 1$K and 7.9\% at $S \ge 4$K}---while epilogue warps scatter outputs. \rev{Offloading scatter to idle communication warps incurs handoff overhead that hurts at small $S$ ($S = \text{batch size} \times \text{sequence length}$) but is negligible at large $S$: in Fig.~\ref{fig:warprole}b, epilogue scatter wins at $S$=128 whereas comm scatter wins at 20K, so no static choice suffices.} \rev{In the \emph{tail} phase, MegaMoE enforces an all-GPU barrier before top-$k$ reduce, stalling faster GPUs on stragglers; even afterward, only epilogue warps execute reduce while all other warps idle, wasting another 4.1\%--6.2\%. Overall (Fig.~\ref{fig:warprole}e), the three bubbles waste 32.8\%--34.1\% of warp-time at $S \le 1$K, 25.1\% at 4K, 21.2\% at 8K, and 15.8\% at 20K.}

\subsubsection{\rev{Selecting Spatial and Temporal Schedules at Runtime}}

\looseness=-1 Policy selection depends on routing statistics available only at runtime, must evaluate \rev{spatial and temporal schedules} jointly because each affects the other's progress, and must remain fast enough not to offset scheduling gains, motivating a lightweight on-device cost model.

\section{Design}\label{sec:design}\label{sec:overview}

\begin{figure}[t]
\centering
\setlength{\unitlength}{1in}
\begin{picture}(3.48,1.56)\put(0,0){\includegraphics{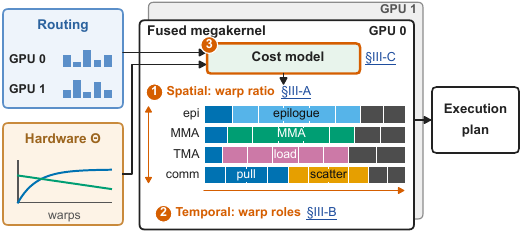}}\input{figures/fig-design-overview-links.tex}\end{picture}
\caption{System design overview of \sys.}
\label{fig:design}
\end{figure}

\rev{As Fig.~\ref{fig:design} shows, \sys combines routing statistics and calibrated hardware parameters to select each GPU's \emph{spatial} and \emph{temporal schedules} on device at launch.}

\subsection{Spatial Scheduling: the Warp Ratio}\label{sec:graph}

\rev{Spatially, \sys partitions warps between communication (\texttt{comm}) and compute/epilogue (\texttt{epi}) per GPU according to runtime demand. Unlike SM-splitting designs that leave communication SMs' matrix units idle, \sys keeps Tensor or Matrix Core control warps active on every SM or CU and shifts only remaining warps between roles. On SM103, hardware constraints restrict per-SM \texttt{comm} and \texttt{epi} counts to multiples of four (Fig.~\ref{fig:rolegraph}); SM90 and CDNA3 likewise expose discrete per-SM/CU ratios (summarized in Table~\ref{tab:spatial}). The trade-off is architecture-dependent: on SM103, dedicated TMEM decouples MMA from epilogue writeback so extra \texttt{comm} warps mainly trade off \texttt{epi} bandwidth, whereas on SM90 and CDNA3 shared accumulators mean extra \texttt{comm} warps reduce the math warps or waves available for GEMMs.}

\begin{table}[t]
\centering
\scriptsize
\caption{Spatial warp/wave design space across evaluated GPU architectures.}
\label{tab:spatial}
\setlength{\tabcolsep}{2.5pt}
\begin{tabular}{@{}lllll@{}}
\toprule
\textbf{GPU} & \textbf{Per SM/CU} & \textbf{Prec.} & \textbf{(\texttt{comm}, comp/epi)} & \textbf{Trade-off} \\
\midrule
SM103 & 16 warps & FP8/FP4 & $(4,8)$, $(8,4)$ \texttt{epi} & TMEM (\texttt{epi} BW) \\
SM90 & 20 warps & FP8/BF16 & $(2,16)$, $(10,8)$ math & Shared regs (math) \\
CDNA3 & 8 waves & FP8/BF16 & $(0,8)$, $(4,4)$ math & Shared regs (math) \\
\bottomrule
\end{tabular}
\end{table}

\looseness=-1 \rev{Across SMs, \sys allows different SMs on one GPU to run different warp ratios: each persistent CTA compares its hardware SM ID against a cost-model threshold to pick between two intra-SM ratios, finely controlling GPU-wide communication warps $N_c$ without multiplying kernel variants while keeping all matrix units active. Across GPUs, peer-to-peer queues decouple sender and receiver warp counts, allowing each GPU to choose its own spatial schedule $r$ independently from local routing statistics to match the per-GPU skew in Fig.~\ref{fig:warprole}c.}

\subsection{Temporal Scheduling: Dynamic Warp Roles}\label{sec:time}

\begin{figure}[t]
\centering
\includegraphics[width=0.92\columnwidth]{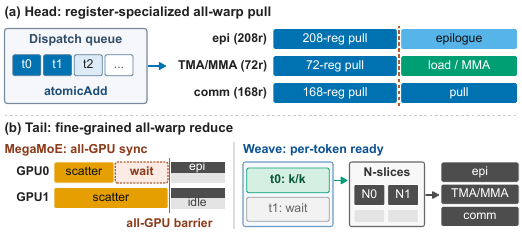}
\caption{Temporal scheduling in \sys: (a) register-specialized all-warp head pull; (b) fine-grained reduce vs.\ MegaMoE's all-GPU synchronization.}
\label{fig:temporal}
\end{figure}

\looseness=-1 Temporally, \sys eliminates Fig.~\ref{fig:rolegraph}'s three bubbles via dynamic warp transitions and fine-grained overlap. Following~\cite{deepgemm,mpk,eventtensor}, warps claim tasks from a global queue via \texttt{atomicAdd}. In the \emph{head} phase (Fig.~\ref{fig:temporal}a), all warps initially pull remote inputs; because \texttt{setmaxnreg} assigns distinct register budgets (72, 168, and 208 regs/thread for \texttt{TMA}/\texttt{MMA}, \texttt{comm}, and \texttt{epi} on SM103), \sys compiles a separate pull routine per budget so warps with more registers keep more loads in flight without spilling. After a preset token count, \texttt{TMA}, \texttt{MMA}, and \texttt{epi} warps switch to GEMMs while \texttt{comm} warps finish pulling.

In the \emph{middle} phase (Fig.~\ref{fig:rolegraph}), \sys selects between \emph{epilogue scatter} ($\mathrm{epi}$), writing outputs directly without handoff overhead, and \emph{comm scatter} ($\mathrm{comm}$), where freed \texttt{comm} warps claim ready outputs via the global queue and stream them through shared memory under ongoing GEMMs. In the \emph{tail} phase (Fig.~\ref{fig:temporal}b), instead of an all-GPU barrier, \sys tracks per-token arrival counts locally and begins reducing each token once its $k$ expert outputs arrive. Every warp group (\texttt{epi}, \texttt{TMA}/\texttt{MMA}, \texttt{comm}) joins reduce via the global queue once free, and \sys slices each token along hidden dimension $N$ to keep all warps busy.

\begin{table}[t]
\centering
\scriptsize
\caption{Notation for the per-GPU cost model.}
\label{tab:theta}
\setlength{\tabcolsep}{3pt}
\begin{tabularx}{\columnwidth}{@{}l >{\raggedright\arraybackslash}X@{}}
\toprule
\textbf{Symbol} & \textbf{Meaning} \\
\midrule
$r,\, s$ & Spatial warp schedule; scatter path ($s \in \{\mathrm{epi}, \mathrm{comm}\}$) \\
$N_c,\, N_e,\, B_M(r)$ & GPU-wide \texttt{comm} and \texttt{epi} warps; token-tile height \\
$m_e,\, M,\, D_{\mathrm{in}},\, M_{\mathrm{out}}$ & Tokens of local expert $e$, $M\!=\!\sum_e m_e$; pull/scatter counts \\
$H,\, I;\, b_w,\, b_a,\, b_o$ & Model dims; weight, GEMM-act., and output byte widths \\
$B_{\mathrm{link}},\, b_{\mathrm{pull}},\, b_{\mathrm{scat}}$ & Unidirectional theoretical peak link BW; per-warp BWs \\
$B_{\mathrm{mem}},\, P_{\mathrm{tc}}(r)$ & Calibrated HBM BW and Tensor/Matrix Core throughput \\
$B_{\mathrm{epi}}(N_e),\, T_{\mathrm{hoff}}$ & Calibrated \texttt{epi} writeback BW; \texttt{comm}-scatter sync overhead \\
\bottomrule
\end{tabularx}
\end{table}

\subsection{Cost Model and Policy Selection}\label{sec:select}\label{sec:cost}

At launch on each GPU, CTA~0 reads routing counts ($m_e$, $M\!=\!\sum_e m_e$, $D_{\mathrm{in}}$, $M_{\mathrm{out}}$) to score each candidate $(r, s)$ using Table~\ref{tab:theta}. Since link bandwidth scales linearly with warps up to $B_{\mathrm{link}}$ (Fig.~\ref{fig:warprole}a, d), pull and scatter take $T_{\mathrm{pull}}(N_c) = D_{\mathrm{in}} H b_a / \min(B_{\mathrm{link}}, N_c b_{\mathrm{pull}})$ and $T_{\mathrm{scat}}(N) = M_{\mathrm{out}} H b_o / \min(B_{\mathrm{link}}, N b_{\mathrm{scat}})$.

Each local expert $e$ holds $3HI b_w$ weight bytes across GEMM0 ($H\!\times\!2I$) and GEMM1 ($I\!\times\!H$) at $6HI$ FLOPs/row. Under tile height $B_M(r)$, its $m_e$ tokens form $K_e(r) = \lceil m_e / B_M(r) \rceil$ tiles padded to $B_M(r)$ rows, loading weights once per tile while $M$ pairs read $H$- and $I$-wide activations:
\begin{equation}
{\medmuskip=1.2mu\thickmuskip=1.8mu\begin{aligned}
F_{\mathrm{gemm}}(r) &= {\textstyle\sum_e} K_e(r) B_M(r) \cdot 6 H I,\\[1pt]
V_{\mathrm{mem}}(r)  &= {\textstyle\sum_e} K_e(r) \cdot 3 H I b_w + M b_a(H\!+\!I)\rlap{.}
\end{aligned}}
\label{eq:work}
\end{equation}
Mainloop time is $T_{\mathrm{mma}}(r) = \max(V_{\mathrm{mem}}(r)/B_{\mathrm{mem}},\allowbreak\, F_{\mathrm{gemm}}(r)/P_{\mathrm{tc}}(r))$, and epilogues write $V_{\mathrm{epi}} = M(b_a I + b_o H)$ bytes in $T_{\mathrm{epi}}(N_e) = V_{\mathrm{epi}} / B_{\mathrm{epi}}(N_e)$, summing per-SM rates under mixed ratios. On SM103, TMEM overlaps MMA with epilogue drain so $T_{\mathrm{comp}}(r) = \max(T_{\mathrm{mma}}(r), T_{\mathrm{epi}}(N_e))$; on SM90 and CDNA3, shared registers serialize them into $T_{\mathrm{comp}}(r) = T_{\mathrm{mma}}(r) + T_{\mathrm{epi}}(N_e)$, while larger $N_c$ forces smaller $B_M(r)$ that raises $K_e(r)$ and lowers $P_{\mathrm{tc}}(r)$.

Temporally, $\mathrm{epi}$ scatter sends outputs via $N_e$ \texttt{epi} warps while pulls overlap compute, whereas $\mathrm{comm}$ scatter pays sync overhead $T_{\mathrm{hoff}}$ for $N_c$ \texttt{comm} warps to pull then scatter:
\begin{equation}
{\medmuskip=0.8mu\thickmuskip=1.2mu\begin{aligned}
T(r,\!\mathrm{epi})  &= \max(T_{\mathrm{pull}}(N_c),\, T_{\mathrm{comp}}(r)\!+\!T_{\mathrm{scat}}(N_e)),\\[1pt]
T(r,\!\mathrm{comm}) &= \max(T_{\mathrm{pull}}(N_c)\!+\!T_{\mathrm{scat}}(N_c),\, T_{\mathrm{comp}}(r)\!+\!T_{\mathrm{hoff}})\rlap{.}
\end{aligned}}
\label{eq:scatter-policy}
\end{equation}
CTA~0 selects $\arg\min_{(r,s)} T(r, s)$ from Eq.~\eqref{eq:scatter-policy} on device.

\subsection{Implementation}\label{sec:impl}

\rev{\sys is implemented in CUDA C++ on SM103/SM90 and HIP C++ on CDNA3 using the per-SM/CU configurations in Table~\ref{tab:spatial}. On SM103, 4 warps handle TMA, MMA, and scheduling while remaining warps mix $4/8$ or $8/4$ (\texttt{comm}, \texttt{epi}) across SMs; on SM90, a producer warpgroup (2 TMA and 2 \texttt{comm} warps) pairs with 4 math/\texttt{comm} warpgroups, and both NVIDIA GPUs use \texttt{setmaxnreg} to resize registers. On CDNA3 (FP32 reduce), CUs run $(4,4)$ or $(0,8)$ (\texttt{comm}, math) waves over inter-node GDA RDMA and intra-node xGMI under two-node EP16.}

\looseness=-1 \rev{Following~\cite{deepgemm,deepep,comet}, \sys adopts three standard optimizations: (1) \texttt{comm} warps pull remote tokens into HBM through registers rather than shared memory, preserving multi-stage GEMM buffering; (2) with cross-GPU deduplication~\cite{deepep,comet}, the worklist prioritizes local tokens ahead of remote pulls to hide initial latency; and (3) \sys allows each expert to pick its own block size $BM_e \in \{32, 64, 128, 256\}$, choosing as large a block as possible while minimizing padding.}

\section{Evaluation}\label{sec:eval}

\begin{figure*}[t]
\centering
\includegraphics[width=0.92\textwidth]{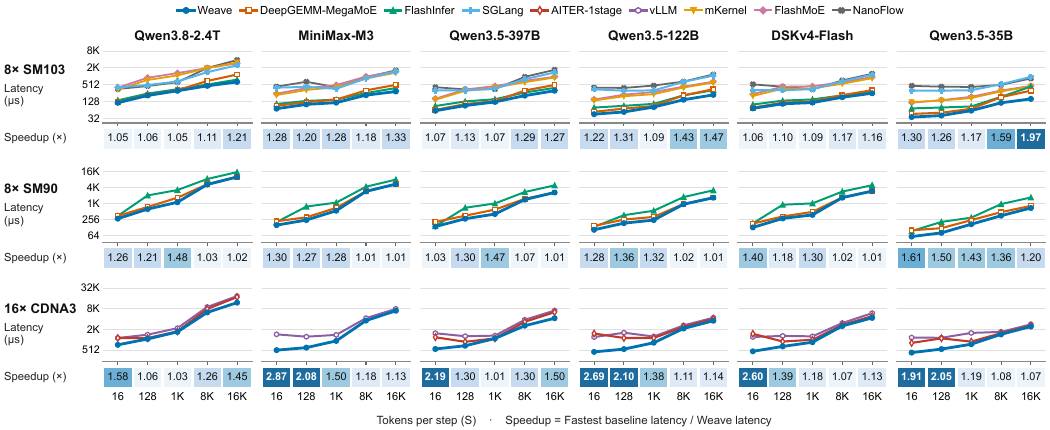}
\caption{MoE-layer latency under real routing on 8\x SM103 (top, EP8), 8\x SM90 (middle, EP8), and 16\x CDNA3 (bottom, 2-node EP16) across six models and five token counts. Lines show latency on a base-2 log scale; shaded cells give \sys speedup over the fastest baseline per workload.}
\label{fig:results}
\end{figure*}

\subsection{Experimental Setup}

\looseness=-1 \textbf{Testbed.} We evaluate on 8\x SM103 (148 SMs, 2032 MHz; CUDA 13.2, PyTorch 2.11.0), 8\x SM90 (78 SMs, 1830 MHz; CUDA 13.3, PyTorch 2.12.1) with NVLink and driver 580.105.08 (EP8), and a two-node 16\x CDNA3 cluster (80 CUs/GPU, 8 xGMI-connected GPUs/node, EP16; ROCm 7.2.0, driver 6.16.13).

\looseness=-1 \textbf{Workloads.} We run Table~\ref{tab:models}'s MoE layers under real ShareGPT routing and randn weights/activations (FP8/NVFP4 on SM103; FP8 on SM90/CDNA3). $S = \text{batch size} \times \text{sequence length}$ spans decoding to prefill with empirical initial pull counts and tail $N$-slices; we report median slowest-GPU latency after warmup.

\looseness=-1 \textbf{Baselines.} On SM103 we compare against DeepGEMM-MegaMoE~\cite{deepgemm} (SM100 bdea79976f, deep\_gemm 2.6.1), FlashInfer~\cite{flashinfer}, SGLang~\cite{sglang} (DeepEP~\cite{deepep}+DeepGEMM), Graph+PDL (five-kernel DeepEP/DeepGEMM in one CUDA Graph with PDL), and tuned SM103 re-implementations of mKernel~\cite{mkernel}, FlashMoE~\cite{flashdmoe} (\texttt{tcgen05}/TMA), and NanoFlow~\cite{nanoflow} (two-stream NCCL/DeepGEMM nano-batches). On SM90 we compare against FP8 MegaMoE (\texttt{nv\_dev} b64107f2) and FlashInfer v0.6.8; on 16\x CDNA3, both official RCCL+AITER \texttt{fused\_moe} paths: one-stage ASM (\texttt{AITER-1stage}, verified on five models; fails FP64 check on MiniMax-M3) and two-stage CK (vLLM~\cite{vllm}); SGLang+MORI exceeds this cluster's 64\,MiB RDMA NIC MR limit.

\begin{table}[t]
\centering
\scriptsize
\caption{MoE configurations of the evaluated models}
\label{tab:models}
\setlength{\tabcolsep}{5pt}
\begin{tabular}{@{}l r r r r@{}}
\toprule
\textbf{Model} & $E$ & $k$ & $H$ & $I$ \\
\midrule
Qwen3.8-2.4T~\cite{qwen38} & 512 & 10 & 8192 & 2048 \\
MiniMax-M3~\cite{minimaxm3} & 128 & 4 & 6144 & 3072 \\
Qwen3.5-397B~\cite{qwen35} & 512 & 10 & 4096 & 1024 \\
Qwen3.5-122B~\cite{qwen35} & 256 & 8 & 3072 & 1024 \\
Qwen3.5-35B~\cite{qwen35} & 256 & 8 & 2048 & 512 \\
DSKv4-Flash~\cite{deepseekv4} & 256 & 6 & 4096 & 2048 \\
\bottomrule
\end{tabular}
\end{table}

\begin{table}[t]
\centering
\scriptsize
\caption{\rev{End-to-end latency (ms) on 8\x SM90 and speedup of \sys over DeepGEMM-MegaMoE}}
\label{tab:e2e}
\setlength{\tabcolsep}{2.0pt}
\begin{tabular}{@{}l rrr rrr@{}}
\toprule
 & \multicolumn{3}{c}{decode ($S$ = 128)} & \multicolumn{3}{c}{prefill ($S$ = 16K)} \\
\cmidrule(lr){2-4}\cmidrule(l){5-7}
Model & MegaMoE & \sys & Speedup & MegaMoE & \sys & Speedup \\
\midrule
Qwen3.8-2.4T & 1.40 & 1.28 & 1.09 & 19.84 & 19.76 & 1.00 \\
Qwen3.5-397B & 0.86 & 0.81 & 1.07 & 5.60 & 5.49 & 1.02 \\
Qwen3.5-122B & 0.65 & 0.60 & 1.08 & 4.05 & 3.99 & 1.02 \\
MiniMax-M3 & 0.91 & 0.81 & 1.12 & 9.31 & 9.36 & 1.00 \\
DSKv4-Flash & 0.90 & 0.87 & 1.04 & 10.11 & 10.10 & 1.00 \\
\midrule
Geomean & & & 1.08 & & & 1.01 \\
\bottomrule
\end{tabular}
\end{table}

\subsection{MoE-Layer Performance}\label{sec:perf}

\looseness=-1 \textbf{Versus MegaMoE and FlashInfer on SM103.} Across the 30 workloads in Fig.~\ref{fig:results} (top), \sys beats DeepGEMM-MegaMoE on every one (1.05--1.97\x, geomean 1.30\x), FlashInfer by 1.11--2.76\x (1.40\x), and the per-workload fastest baseline by 1.05--1.97\x (1.22\x). Speedup grows with $S$: small $S$ is HBM/latency-bound so static epilogue scatter suffices, whereas at $S \ge 8$K (12 workloads) MegaMoE stalls on epilogue scatter while \sys streams scatter via communication warps under GEMMs, achieving 1.18--1.97\x (1.55\x geomean).

\looseness=-1 \textbf{Versus other baselines.} \sys beats SGLang (2.15--8.81\x, geomean 4.28\x), mKernel (2.48--4.96\x, 3.26\x), FlashMoE (2.69--5.29\x, 3.62\x), and NanoFlow (2.03--13.04\x, 5.00\x) on all 30 workloads. NanoFlow's nano-batches repeatedly read expert weights with a static SM split, while mKernel and FlashMoE dedicate whole SMs to communication and leave their Tensor Cores idle; none overlaps communication and compute at warp granularity. \rev{\sys also beats Graph+PDL by 1.82--3.00\x (2.30\x geomean); disabling PDL changes latency by only $-$0.5\% on average because CUDA Graph and PDL only hide launch overhead and prologues, whereas overlapping separate kernels across streams can only partition at SM granularity, wasting the compute of communication SMs.}

\looseness=-1 \textbf{SM90 and CDNA3.} Across the 30 workloads in Fig.~\ref{fig:results} (middle), \sys beats MegaMoE on every one (\rev{1.01--1.72\x, geomean 1.24\x: 1.37\x for $S \le 128$, 1.38\x at $S$=1K, 1.06\x for $S \ge 8$K}), \rev{FlashInfer by 1.03--3.28\x (1.95\x)}, and the fastest baseline by 1.01--1.61\x (1.21\x). Unlike SM103, SM90 gains more at $S \le 1$K than $S \ge 8$K due to its lower compute-to-NVLink ratio (0.92 vs.\ 12.6~TFLOPS per GB/s): at $S \ge 8$K, SM103 is NVLink-bound (1.55\x gain), whereas SM90 is compute-bound where MegaMoE already reaches up to 85.0\% Tensor Core utilization (1.06\x). At $S \le 1$K, SM90 Tensor Cores are underutilized, letting \sys remove warp-scheduling bubbles for 1.37--1.38\x speedup. \rev{On 16\x CDNA3 (EP16, Fig.~\ref{fig:results} bottom), \sys beats every official baseline on all 30 workloads: 1.01--3.41\x (geomean 1.43\x) over \texttt{AITER-1stage} (25 verified cells), 1.13--2.99\x (1.67\x) over two-stage vLLM, and 1.01--2.87\x (1.44\x: 1.91\x for $S \le 128$, 1.20\x at 1K, 1.19\x for $S \ge 8$K) over the per-workload fastest baseline. Unlike MegaMoE on SM103/SM90, RCCL+AITER serializes dispatch, \texttt{fused\_moe}, and combine kernels; fusing the layer and overlapping inter-node GDA RDMA and intra-node xGMI with 8-wave/CU Matrix Core GEMMs and cooperative tail reduction sustains 1.19\x (up to 1.50\x) even at $S \ge 8$K.}

\begin{table}[!t]
\centering
\scriptsize
\caption{Progressive ablation of \sys in geomean speedup over baseline ($-$S: no spatial ratio; $-$S$-$T: also no temporal overlap/tail-reduce; $-$S$-$T$-$C: also no cost model)}
\label{tab:ablation}
\setlength{\tabcolsep}{2.8pt}
\begin{tabular}{@{}ll cccc@{}}
\toprule
Platform & $S$ range & \sys & $-$S & $-$S$-$T & $-$S$-$T$-$C \\
\midrule
SM103 & $\le$128 (12) & \rev{1.165} & \rev{1.076} & \rev{0.915} & \rev{0.913} \\
 & 1K--4K (18) & \rev{1.112} & \rev{1.073} & \rev{0.919} & \rev{0.916} \\
 & $\ge$8K (18) & \rev{1.638} & \rev{1.416} & \rev{1.185} & \rev{1.148} \\
 & All (48) & \rev{1.301} & \rev{1.192} & \rev{1.010} & \rev{0.996} \\
\midrule
SM90 & $\le$128 (12) & \rev{1.255} & \rev{1.226} & \rev{1.133} & \rev{1.010} \\
 & 1K--4K (18) & \rev{1.189} & \rev{1.188} & \rev{1.043} & \rev{1.028} \\
 & $\ge$8K (18) & \rev{1.079} & \rev{1.034} & \rev{1.012} & \rev{0.836} \\
 & All (48) & \rev{1.162} & \rev{1.137} & \rev{1.053} & \rev{0.947} \\
\midrule
CDNA3 & $\le$128 (12) & \rev{1.845} & \rev{1.778} & \rev{1.758} & \rev{1.548} \\
 & 1K--4K (18) & \rev{1.160} & \rev{1.055} & \rev{0.922} & \rev{0.753} \\
 & $\ge$8K (18) & \rev{1.176} & \rev{0.973} & \rev{0.811} & \rev{0.638} \\
 & All (48) & \rev{1.309} & \rev{1.166} & \rev{1.033} & \rev{0.848} \\
\bottomrule
\end{tabular}
\end{table}

\subsection{End-to-End Performance}\label{sec:e2e}

On 8\x SM90 (Table~\ref{tab:e2e}), \sys speeds up decode ($S$=128, 4K context) by \rev{1.04--1.12\x (1.08\x geomean)} and prefill ($S$=16K) by up to \rev{1.02\x (1.01\x geomean)} over MegaMoE. Decode gains more because MoE takes \rev{59\%--75\%} of decode latency, whereas attention takes \rev{40\%--67\%} of prefill as SM90 Tensor Cores saturate.

\subsection{Ablation}\label{sec:ablation}

\looseness=-1 \rev{Table~\ref{tab:ablation} ablates \sys across 48 workloads per platform: $-$S fixes the warp ratio, $-$S$-$T also disables temporal overlap and tail-reduce, and $-$S$-$T$-$C removes the cost model. Geomean speedup falls monotonically from 1.301\x to 1.192\x, 1.010\x, and 0.996\x on SM103 (1.638\x$\to$1.416\x$\to$1.185\x$\to$1.148\x at $S\!\ge\!8$K), from 1.162\x to 1.137\x, 1.053\x, and 0.947\x on SM90, and from 1.309\x to 1.166\x, 1.033\x, and 0.848\x on 16\x CDNA3. On SM103, temporal switching ($-$S vs.\ $-$S$-$T) removes head/tail bubbles at $S\!\le\!128$ (1.076\x vs.\ 0.915\x) and overlaps \texttt{comm} scatter at $S\!\ge\!8$K (1.416\x vs.\ 1.185\x), while spatial tuning adds 15.7\% at $S\!\ge\!8$K. On SM90, temporal scheduling ($-$S vs.\ $-$S$-$T) drives small/medium-$S$ gains (1.226\x vs.\ 1.133\x at $S\!\le\!128$, 1.188\x vs.\ 1.043\x at 1K--4K), while removing the cost model ($-$S$-$T$-$C) drops $S\!\ge\!8$K to 0.836\x as static defaults underperform MegaMoE. On CDNA3, spatial (1.309\x vs.\ 1.166\x), temporal (1.166\x vs.\ 1.033\x), and cost-model selection (1.033\x vs.\ 0.848\x) each contribute across all $S$, lifting $S\!\ge\!8$K from 0.638\x under static defaults to 1.176\x.}

\subsection{Cost-Model Accuracy and Overhead}\label{sec:planner}

\begin{figure}[t]
\centering
\includegraphics[width=0.78\columnwidth]{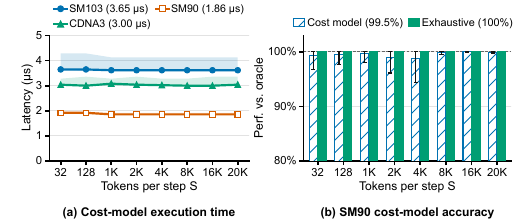}
\caption{On-device cost-model execution time (a) and SM90 cost-model accuracy (b) across 48 workloads.}
\label{fig:planner}
\end{figure}

\looseness=-1 On SM103, offline-selected fixed configurations achieve \rev{1.301\x} geomean over MegaMoE across Table~\ref{tab:ablation}'s 48 workloads versus a \rev{1.332\x} exhaustive oracle, with \rev{2.6\%} mean (2.4\% geomean) \emph{selection error} and \rev{38/48} exact matches. On SM90 (Fig.~\ref{fig:planner}b), fixed selections reach \rev{99.5\%} of oracle (\rev{1.164\x vs.\ 1.170\x}), with 32/48 exact and 41/48 within 1\% error, while in-kernel selection reaches \rev{99.3\%} (\rev{1.162\x}).

\looseness=-1 \textbf{Decision overhead.} In Fig.~\ref{fig:planner}a, CTA-0 cost-model execution takes 3.65\,$\mu$s median on SM103 (3.58--4.29\,$\mu$s), 1.86\,$\mu$s on SM90 (1.86--1.92\,$\mu$s), and 3.00\,$\mu$s on CDNA3 (3.00--3.36\,$\mu$s), independent of $S$. Because CTA-0 scans only local experts and a few candidate policies, latency stays flat across $S$=32--20K on all three GPUs. End-to-end overhead is hidden on SM103 ($-$0.08\,$\mu$s, $\pm$1.3\% paired) and adds \rev{0.6\,$\mu$s median (0.1\%)} on SM90.

\subsection{Combining with EPLB}\label{sec:eplb}

\begin{figure}[t]
\centering
\includegraphics[width=0.74\columnwidth]{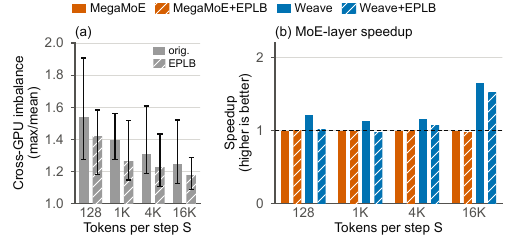}
\caption{Combining with EPLB on SM103: (a) cross-GPU imbalance; (b) geomean speedup over original-routing MegaMoE across six models.}
\label{fig:eplb}
\end{figure}

\looseness=-1 \sys is orthogonal to EPLB~\cite{eplb}: EPLB replicates hot experts to balance tokens across GPUs, whereas \sys schedules warps within each GPU to match communication and compute demands. With 32 redundant experts across 24 workloads (6 models \x $S \in \{128, 1\mathrm{K}, 4\mathrm{K}, 16\mathrm{K}\}$), EPLB cuts max/mean imbalance from 1.37 to 1.27 (Fig.~\ref{fig:eplb}) while adding 12.9\% average latency to \sys; \sys stays 1.274\x faster than MegaMoE on original routing (24/24) and 1.128\x faster with EPLB (17/24). Replicating 32 experts spreads tokens thinner, reducing $m_e$ and raising tile padding and weight loads ($K_e(r)$ in Eq.~\eqref{eq:work}) at small/medium $S$ on MegaMoE (0.986\x) and \sys; at $S$=16K padding amortizes and \sys{}+EPLB reaches 1.52\x over MegaMoE (vs.\ 1.65\x without EPLB).

\section{Related Work and Limitations}\label{sec:related}

\looseness=-1 \textbf{Communication--computation overlap in MoE.} Prior MoE systems~\cite{fastermoe,dsmoe,tutel,smartmoe,lina,janus,lancet,nanoflow} overlap all-to-all transfers with GEMMs across separate kernels or streams via sub-batching, incurring cross-kernel launch, synchronization, and repeated expert-weight read overheads. Building on fine-grained overlap~\cite{t3,flux}, Comet~\cite{comet}, Triton-distributed~\cite{tdist}, FlashMoE~\cite{flashdmoe}, and UniEP~\cite{uniep} pipeline transfers with GEMMs inside fused kernels via on-device synchronization, yet like DeepEP~\cite{deepep} partition thread blocks or SMs statically at compile time; mKernel~\cite{mkernel} adjusts SM counts at runtime, but dedicating whole SMs to communication still idles their Tensor Cores.

\looseness=-1 \textbf{Intra-SM warp specialization.} Kernel libraries~\cite{cutlass,fa3,tk} specialize warps within each SM between asynchronous data movement and Tensor Core math, and compilers~\cite{cypress,tawa,twill} automate warp-specialized software pipelining for single-GPU kernels. For distributed MoE, MegaMoE~\cite{deepgemm} co-locates communication and GEMM warps on every SM, keeping all Tensor Cores active, yet fixes warp counts and roles at compile time. In contrast, \sys schedules warps spatially and temporally at runtime via an on-device cost model.

\looseness=-1 \textbf{Limitations.}\label{sec:discussion} \sys targets NVIDIA and AMD GPUs and has not yet been evaluated on non-GPU accelerators whose compute-core and DMA hierarchies differ from GPU warps and wavefronts.

\section{Conclusion}

\looseness=-1 We presented \sys, a routing-adaptive MoE megakernel that schedules warps spatially and temporally at runtime to match per-GPU loads under EP. Combining SM- and warp-level spatial partitioning, phase-aware warp-role switching, and a 1.86--3.65\,$\mu$s on-device cost model, \sys outperforms DeepGEMM-MegaMoE by 1.30\x geomean (up to 1.97\x) on 8\x SM103 and 1.24\x (up to 1.72\x) on 8\x SM90, and beats the fastest official RCCL+AITER baseline by \rev{1.44\x} (up to \rev{2.87\x}; \rev{1.67\x} over vLLM) on 16\x CDNA3. It speeds up 8\x SM90 end-to-end decoding by 1.08\x geomean, reaches within 2.6\% of oracle latency on SM103 and 99.5\% on SM90 with negligible overhead, and complements EPLB. Thus, fine-grained, routing-adaptive warp scheduling unlocks sustained communication--computation overlap across diverse GPU architectures.

\clearpage
\bibliographystyle{IEEEtran}
\bibliography{refs}

\end{document}